\documentclass[twocolumn,showpacs,preprintnumbers,amsmath,amssymb,aps,bibnotes,superscriptaddress]{revtex4}

\usepackage{amsfonts}
\usepackage[dvips]{graphicx}
\usepackage[ps2pdf,bookmarks=true,bookmarksnumbered=true,breaklinks=true,pdfborder={0 0 112.0}]{hyperref}
\usepackage{color}
\usepackage{bm}

\newcommand{\eqnref}[1]{Eq.~\eqref{#1}}
\newcommand{\figref}[1]{Fig.~\ref{#1}}

\newcommand{\ket}[1]{|#1\rangle}
\newcommand{\e}[1]{\text{e}^{#1}}
\newcommand{\cmplxi}{\text{i}}

\renewcommand{\vec}[1]{\mathbf{#1}}

\newcommand{\punc}[1]{\,#1}

\newcommand{\neweqnline}{\nonumber\\}
\newcommand{\diffd}{\text{d}}

\def \s{{\sigma}}

\begin{document}
\title{Effect of three-body loss on itinerant ferromagnetism in an atomic Fermi gas}
\author{G.J.~Conduit}
\email{gjc29@cam.ac.uk}
\affiliation{Department of Condensed Matter Physics, Weizmann Institute of Science, Rehovot, 76100, Israel}
\affiliation{Physics Department, Ben Gurion University, Beer Sheva, 84105, Israel}
\author{E.~Altman}
\affiliation{Department of Condensed Matter Physics, Weizmann Institute of Science, Rehovot, 76100, Israel}
\date{\today}

\begin{abstract}
A recent experiment has provided the first evidence for itinerant
ferromagnetism in an ultracold atomic gas of fermions with repulsive
interactions. However, the gas in this regime is also subject to
significant three-body loss. We adopt an extended Hertz-Millis
theory to account for the effect of loss on the transition and on the
ferromagnetic state.  We find that the losses damp quantum
fluctuations and thereby significantly increase the critical
interaction strength needed to induce ferromagnetism. This effect may
resolve a discrepancy between the experiment and previous theoretical
predictions of the critical interaction strength.  We further
illuminate the impact of loss by studying the collective spin
excitations in the ferromagnet. Even in the fully polarized state,
where loss is completely suppressed, spin waves acquire a decay rate
proportional to the three-body loss coefficient.
\end{abstract}


\maketitle

The Stoner transition from a paramagnetic metal to a ferromagnet is one
of the earliest known and seemingly simple examples of a quantum phase
transition. Yet recent theoretical work~\cite{Belitz98,Conduit09} has
revealed a great deal of complexity and suggested that quantum
fluctuations play a vital role in determining the behavior near to the
quantum critical point. Specifically fluctuations drive the
ferromagnetic transition first order at low temperature and may lead
to the formation of novel phases~\cite{Conduit09}. Whether these
effects can explain puzzling experimental observations in materials
such as $\text{ZrZn}_{2}$ and
$\text{Sr}_{3}\text{Ru}_{2}\text{O}_{7}$~\cite{Borzi07}, or whether
coupling to phonons or other auxiliary degrees of freedom is involved
remains an open question.  Ultracold Fermi gases tuned by a Feshbach
resonance now offer experimentalists unprecedented control over
many-body phenomena and so provide a concrete platform from which to
answer such questions and enhance our understanding of quantum
critical phenomena in itinerant
ferromagnets~\cite{Coleman08,LeBlanc09,Conduit08}.

However, the atomic systems are not free of complications of their
own.  A major obstacle to the formation of a ferromagnetic phase in an
ultracold atomic gas is the loss of atoms due to three-body
interactions~\cite{Petrov03}. A recent and seminal
experiment~\cite{Jo09} overcame this hurdle with a non-adiabatic
Feshbach field quench to festinate into the ferromagnetic state, and
has now provided the first firm evidence of the formation of a
ferromagnetic state in an atomic gas~\cite{Conduit10}. The experiment
has suggested that the ferromagnetic transition takes place at
$k_{\text{F}}a\approx2.2$, which is in stark contrast to the Quantum
Monte Carlo prediction of $k_{\text{F}}a\approx0.85$~\cite{Conduit09}
or the analytical prediction with quantum fluctuations of
$k_{\text{F}}a\approx1.05$~\cite{Conduit08}. Although both the trap
geometry and non-adiabatic conditions can artificially raise the
required interaction strength~\cite{Conduit10}, the source of the
discrepancy in the interaction strength has yet to be adequately
understood. It is vital to resolve this inconsistency to definitively
prove that the ferromagnetic phase was formed rather than some
alternative strongly correlated phase~\cite{Zhai09}.

In this letter we investigate how the three-body loss interaction
impacts on the Stoner transition and the behavior of the system in the
ferromagnetic phase.  The loss can be viewed as an additional
interaction that may renormalize the value of the repulsive
interaction, as well as introduce fluctuations of its own that give
rise to new phenomena.  For example, it was recently argued that
strong two or three body loss in a Bose system can give rise to
effective hard core interactions~\cite{Syassen09,Durr09,Daley09}, and
in particular and lead to dynamic formation of a Tonks-Girardeau
gas~\cite{Syassen09}. In the itinerant fermion system, we discover
that three-body losses inhibit quantum fluctuations out of the fully
polarized state which, since quantum fluctuations promote
ferromagnetism~\cite{Conduit08,Conduit09}, raises the interaction
strength required to stabilize the ferromagnetic state. This mechanism
provides strong motivation to study the collective modes of the fully
polarized phase. We find that spin waves acquire a finite life-time in
presence of loss, even in the fully polarized phase where loss is
completely suppressed in the pristine system.

\emph{The renormalization of interactions due to atom loss} can be
studied by adapting the formalism developed in Ref.~\cite{Conduit08}
that demonstrated how quantum fluctuations drive the ferromagnetic
phase transition first order. Since the prediction of the interaction
strength $k_{\text{F}}a=1.054$ for the onset of ferromagnetism is
backed up by robust Quantum Monte Carlo calculations~\cite{Conduit09},
this formalism provides a solid foundation for the present study.

In order to study the effect of loss using a framework suited for
equilibrium systems, we adopt a generalized linear response strategy
in which we seek the modified quasi-equilibrium state established by
the small loss term. To abrogate the three-body loss we insert an
artificial single-body atom source term.  Then following the
prescription laid out in Ref.~\cite{Conduit08} we integrate out
quantum fluctuations and find that the three-body loss renormalizes
the interaction strength, a consequence of atom loss damping the
quantum fluctuations. Once the magnetization is established we focus
on those regions of the phase diagram where atom loss and therefore
the source term is zero and so faithful to the experimental
setup. Having laid out the strategy we now present the formal
calculation including atom loss.

Our goal is to calculate the quantum partition function with a
fermionic coherent state path integral
$\mathcal{Z}=\int\mathcal{D}(\psi,\bar{\psi})\e{-S[\psi,\bar{\psi}]}$. The
starting point is the action $S[\psi,\bar{\psi}]$,
which describes a two component Fermi gas with pseudospin
$\sigma\in\{\uparrow,\downarrow\}$ that is represented by fermionic
fields $\psi$ and $\bar{\psi}$
\begin{align}
S\!=\!\!\int_{0}^{\beta}\!\!\diffd\tau\diffd\vec{r}\!\left[\sum_{\sigma}\bar{\psi}_{\sigma}\left(\partial_{\tau}\!+\!\epsilon_{\vec{k}}\!-\!\mu\right)\psi_{\sigma}\!+\!g\bar{\psi}_{\uparrow}\bar{\psi}_{\downarrow}\psi_{\downarrow}\psi_{\uparrow}\right]\!\!\punc{,}
\end{align}
where the atoms have the single particle dispersion
$\epsilon_{\vec{k}}=k^{2}/2$ and feel a repulsive s-wave contact
interaction $g\delta^{3}(\vec{r})$ that can be tuned via a Feshbach
resonance, and $\beta=1/k_{\text{B}}T$ is the inverse temperature. We
have also set $\hbar=m=1$. Before proceeding we consider how to
incorporate atom loss into the model. If we momentarily consider the
non-interacting system, and insert a hypothetical imaginary term
$\cmplxi\chi\bar{\psi}_{\sigma}\psi_{\sigma}$ into the 
  Lagrangian, then the corresponding retarded Green's function is
$\hat{G}_{\vec{k}}^{+}=(\partial_{\tau}+\epsilon_{\vec{k}}-\mu+\cmplxi\chi)^{-1}$. If
we now perform the analytical continuation $\tau\mapsto\cmplxi t$ into
real time, and consider the Fourier transform to the temporal domain
we recover $G_{\vec{k}}^{+}(t)=\exp[\cmplxi
  t(\epsilon_{\vec{k}}-\mu)-\chi t]\Theta(t)$. Therefore, within
linear response theory, we can identify an imaginary term in the
action with the effective lifetime $\chi^{-1}$ of the state
$\ket{\vec{k}}$, and therefore a probe of atom loss.

In the cold atom gas the dominant loss process is three-body
interactions.  Following the linear response template we incorporate
three-body loss into the action through the term
$\cmplxi\lambda\bar{\rho}\bar{\psi}_{\uparrow}\bar{\psi}_{\downarrow}\psi_{\downarrow}\psi_{\uparrow}$.
Here we replaced a pair of dynamic fermion fields with the average
density ${\bar \rho}$, which amounts to a two-body mean-field
approximation. The three-body loss term is now on equal footing to the
two-body interaction, and so is perfectly poised to study the
renormalization of the interaction strength.  The coefficient
$\lambda$ is estimated, in the weakly interacting regime $k_{\text{F}}a<<1$,
to be $\lambda=111\bar{\epsilon}(k_{\text{F}}a)^{6}$~\cite{Petrov03}.
Since we seek a quasi-equilibrium state, we counterpoise the atom loss
with an atom source term
$-\cmplxi\gamma\sum_{\sigma}\bar{\psi}_{\sigma}\psi_{\sigma}$, where
$\gamma$ will be determined later. Within this formalism, the action
is
\begin{align}
 S&=\int_{0}^{\beta}\diffd\tau\diffd\vec{r}\Biggl[\sum_{\sigma}\bar{\psi}_{\sigma}\left(\partial_{\tau}+\epsilon_{\vec{k}}-\mu-\cmplxi\gamma\right)\psi_{\sigma}\neweqnline
 &+\left(g+\cmplxi\lambda\bar{\rho}\right)\bar{\psi}_{\uparrow}\bar{\psi}_{\downarrow}\psi_{\downarrow}\psi_{\uparrow}\Biggr]\punc{.}
\end{align}

To proceed we calculate the free energy following the prescription
laid out in Ref.~\cite{Conduit08}. First we introduce a
Hubbard-Stratonovich transformation in both the density channel $\rho$
and the magnetization channel ${\bm\phi}$ to decouple the quartic
terms in the fermionic field. This leads us to identify the spectrum
$\xi_{\vec{k},\sigma}=\epsilon_{\vec{k}}+\cmplxi\gamma_{\vec{k}}+(g+\cmplxi\lambda\bar{\rho})(\rho-\sigma\phi)-\mu$.
After integrating out the fermionic variables we expand the
fluctuations in the bosonic fields to quadratic order and also
integrate them out. To remove the unphysical ultraviolet divergence of
the contact interaction we employ the standard regularization setting
$g\mapsto\frac{2k_{\text{F}}a}{\pi\nu}-\frac{2}{V}(\frac{2k_{\text{F}}a}{\pi\nu})^{2}\sum_{\vec{k}_{3,4}}'(\xi_{{\bf
    k}_1,\uparrow}+\xi_{{\bf k}_2,\downarrow}- \xi_{{\bf
    k}_3,\uparrow}-\xi_{{\bf
    k}_4,\downarrow})^{-1}$~\cite{Pathria07}. The prime indicates that
the summation is subject to the momentum conservation ${\bf k}_1+{\bf
  k}_2={\bf k}_3+{\bf k}_4$, the Fermi distribution $n$ is calculated
in the presence of the now imaginary chemical potential, and $\nu$ is
the density of states at the Fermi surface of an equivalent
non-interacting gas. This regularization allows us to characterize the
strength of the interaction through the dimensionless parameter
$k_{\text{F}}a$, where $k_{\text{F}}$ denotes the Fermi wave vector
and $a$ is the s-wave scattering length. The analysis yields a
perturbation expansion in terms of the dimensionless interaction
strength and the loss parameter
\begin{align}
 S&=\sum_{\sigma,\vec{k}}(\epsilon_{\vec{k}}+\cmplxi\gamma_{\vec{k}})n(\xi_{\vec{k},\sigma})+\left[\frac{2k_{\text{F}}a}{\pi\nu}+\cmplxi\lambda\bar{\rho}\right]\rho_{\uparrow}\rho_{\downarrow}\neweqnline
  &-2\left[\frac{2k_{\text{F}}a}{\pi\nu}+\cmplxi\lambda\bar{\rho}\right]^{2}\Upsilon\punc{,}
\end{align}
where the quantum fluctuations are encoded in the term
\begin{align}
 \Upsilon={\sum_{\vec{k}_{1,2,3,4}}}^{\!\!\!'}\frac{n
(\xi_{\vec{k}_{1},\uparrow})n(\xi_{\vec{k}_{2},\downarrow})
\left[n(\xi_{\vec{k}_{3},\uparrow})+n
(\xi_{\vec{k}_{3},\downarrow})\right]}{\xi_{\vec{k}_{1},\uparrow}+\xi_{\vec{k}_{2},\downarrow}
-\xi_{\vec{k}_{3},\uparrow}-\xi_{\vec{k}_{4},\downarrow}}\punc{.}
\end{align}
A similar expression was derived in the homogeneous case in
Ref.~\cite{Abrikosov58} using second order perturbation theory.
Consistent with the initial identification of atom loss through linear
response theory, and the perturbation expansion in the interaction
strength, we now expand the action out to quadratic order in $\gamma$
and $\lambda$, and assume that the gas is at low temperature so that
$n(\xi+\Delta)\approx
n(\xi)-\Delta\delta(\xi)-\Delta^{2}\delta'(\xi)/2$. To establish the
connection to the free energy we first demand that there is no
imaginary component to the free energy which fixes the atom source
term to
\begin{align}
 \gamma=\lambda\bar{\rho}\frac{\rho_{\uparrow}\rho_{\downarrow}-2(\rho_{\uparrow}\mu_{\downarrow}\nu_{\downarrow}+\rho_{\downarrow}\mu_{\uparrow}\nu_{\uparrow})}{\bar{\rho}-\mu_{\uparrow}\nu_{\uparrow}-\mu_{\downarrow}\nu_{\downarrow}}\punc{.}
\label{gamma}
\end{align}
Here $\nu_\s$ denotes the density of states at the Fermi energy of
species $\s$.  We note that the atom loss is zero when the system is
fully polarized. The real part of the action corresponds to the free
energy $F=F_0+\Lambda(\lambda)$ modified by atom loss. The standard
theory of a lossless system is encoded in the term
\begin{align}
 F_0=\sum_{\sigma,\vec{k}}\epsilon_{\vec{k}}n(\xi_{\vec{k},\sigma})+\frac{2k_{\text{F}}a}{\pi\nu}\rho_{\uparrow}\rho_{\downarrow}-2\left(\frac{2k_{\text{F}}a}{\pi\nu}\right)^{2}\Upsilon\punc{.}
\end{align}
The renormalization of the interaction strength due to atom loss
enters through
\begin{align}
 \Lambda&=2\lambda^{2}\bar{\rho}^{2}\left(\Upsilon-\rho_{\uparrow}^{2}\mu_{\downarrow}\nu'_{\downarrow}-\rho_{\downarrow}^{2}\mu_{\uparrow}\nu'_{\uparrow}\right)\neweqnline
 &+\lambda\bar{\rho}\gamma\left[\rho_{\uparrow}(\nu_{\downarrow}-2\mu_{\downarrow}\nu'_{\downarrow})+\rho_{\downarrow}(\nu_{\uparrow}-2\mu_{\uparrow}\nu'_{\uparrow})\right]\neweqnline
 &+\gamma^{2}\left[\frac{\nu_{\uparrow}+\nu_{\downarrow}}{2}-\mu_{\uparrow}\nu'_{\uparrow}-\mu_{\downarrow}\nu'_{\downarrow}\right]\punc{,}
 \label{eqn:AtomLossCoefficient}
\end{align}
where $\nu'_{\sigma}$ is the differential of the density of states at
the Fermi surface of species $\sigma$.
In our further analysis of the free energy it will prove
convenient to consider the dimensionless loss parameter
$\tilde{\lambda}=\pi\nu\lambda\bar{\rho}/2$.

\begin{figure}
 \centerline{\resizebox{0.85\linewidth}{!}{\includegraphics{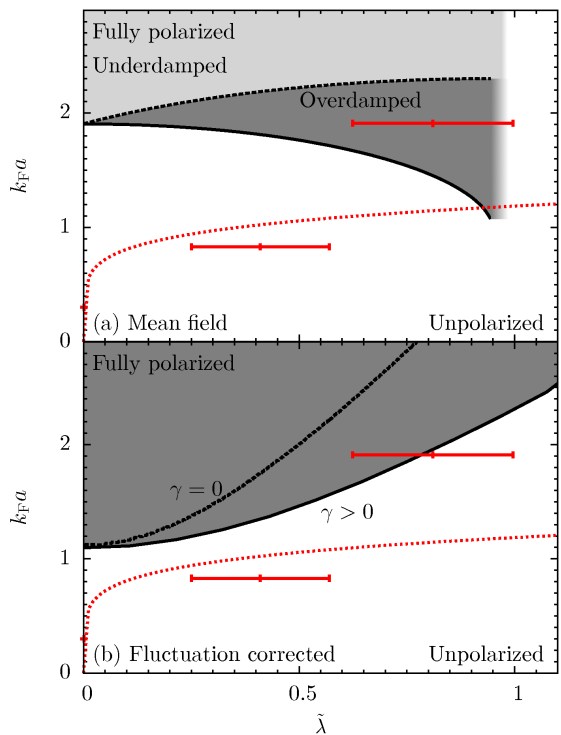}}}
 \caption{(Color online) The required interaction strength to reach
   the shaded full polarization with changing loss rate $\lambda$ in
   (a) the mean-field case and (b) when fluctuation corrections are
   considered. The atom loss variation with weak interactions is
   shown by the red dotted line, and experimental~\cite{Huckans09}
   estimates of atom loss by the red points with error bars. In (a)
   the regions where collective modes are underdamped and overdamped
   are highlighted, in (b) the fully polarized boundary with
   $\gamma>0$ (solid) and $\gamma=0$ (dotted) is plotted.}
 \label{fig:LossyFerroIntVar}
\end{figure}

We now use the formalism to determine the phase diagram, focusing on
the boundary of the fully polarized state. The mean-field case
presented in \figref{fig:LossyFerroIntVar}(a) provides a solid
foundation on which to build the analysis.  Quantum fluctuations
destroy the ferromagnetic state, but atom loss damps fluctuations and
stabilizes the fully polarized state, which consequentially can be
seen at weaker interaction strengths. This behavior stems directly
from \eqnref{eqn:AtomLossCoefficient}, which has a negative gradient
with magnetization. However, Refs.~\cite{Conduit08,Conduit09}
demonstrated that to higher order in the interaction strength quantum
fluctuations in fact stabilize the ferromagnetic phase. Therefore,
when higher order corrections are taken into account in
\figref{fig:LossyFerroIntVar}(b), quantum fluctuations at
$\tilde{\lambda}=0$ reduce the critical interaction strength compared
to the mean-field case. These quantum fluctuations also drive the
transition first order.  With increasing three-body loss the quantum
fluctuations are damped, which in turn raises the interaction strength
required to see ferromagnetism.  This feature is repeated in
\eqnref{eqn:AtomLossCoefficient} which has a positive gradient with
magnetization. Beyond a critical value of the loss coefficient
$\lambda\gtrsim1.7$ which is of the order of the interaction strength
$g=\pi/2$ at the Stoner criterion for the second order transition, the
fluctuations that drove the ferromagnetic transition first order are
sufficiently suppressed to recover that second order transition.  At
high interaction strengths the border adopts a linear behavior as the
highest order terms in the free energy in interaction strength and
loss are both quadratic.  However, in this regime it is likely that
higher order corrections will play an important role.  The curve was
calculated with the atom source bath in place; we also verified the
robustness of our result by making the alternative approximation of
explicitly setting $\gamma=0$ and ignoring the atom loss. In this case
the curve (also shown in \figref{fig:LossyFerroIntVar}b) is found to
be qualitatively the same, though with an enhanced critical
interaction strength.

We now turn to consider the experimental repercussions of the
renormalization of the interaction strength.  The theoretical
prediction \cite{Petrov03} for the variation of the loss coefficient
as the system is tuned closer to the Feshbach resonance is given by
$\lambda=111\bar{\epsilon}(k_{\text{F}}a)^{6}$.  As seen in
\figref{fig:LossyFerroIntVar}(b), this trajectory does not enter the
fully polarized regime, so according to this model the system remains
steadfastly in the paramagnetic (or at most partially polarized)
phase. However this prediction is only valid for $k_{\text{F}}a\ll1$. Indeed
the loss coefficient as estimated from the experiment
~\cite{Huckans09} (points with error bars) is much lower for given
$k_{\text{F}}a$. With this estimate the system in fact passes into the fully
polarized state. Though the error bars on the experimental results are
large, the results indicate that the system becomes fully polarized at
$k_{\text{F}}a\approx1.9$, which compares favorably with the
experimental observation that the atomic gas became fully polarized at
$k_{F}a\approx2.2$~\cite{Jo09}.  The experimental observation of the
raised critical interaction strength is strongly suggestive of the
important role that quantum fluctuations have to play.  The sole focus
on the fully polarized gas will make useful predictions for experiment
since the analysis on the homogeneous system showed that the system is
partially polarized over the narrow range
$1.05<k_{\text{F}}a<1.11$~\cite{Conduit08}, compared with the broad
ambit of interaction strengths that exist within the trap
$0<k_{\text{F}}a<8$~\cite{Conduit10}.  Therefore, the partially
polarized regime not considered here fills only a thin shell within
the trap.

Having exposed the significant renormalization of
interaction strength due to three-body interactions we now demonstrate
that the dispersion of the collective modes is a cogent probe of the
impact of loss driven damping on the quantum fluctuations.

\emph{The dispersion of the collective modes} can be studied by
searching for the poles in the propagators. We search only for
the collective modes in the fully ferromagnetic state where the
longitudinal mode is gapped so we focus on the transverse modes
$(\Omega,\vec{q})$ that have the inverse propagator
$1+\frac{2k_{\text{F}}a}{\pi\nu}\sum_{\omega,\vec{p}}G_{\uparrow}(\omega+\Omega,\vec{p}+\vec{q})G_{\downarrow}(\omega,\vec{p})$~\cite{Conduit08}. Equating this to zero yields collective mode frequencies with both a real and an imaginary
part. The real part, which gives the collective mode dispersion is
\begin{align}
 \Omega=\frac{q^{2}}{2}\left(1-\frac{2^{5/3}3}{5k_{\text{F}}a}\frac{1}{1+\tilde{\lambda}^{2}/(k_{\text{F}}a)^{2}}\right)\punc{.}
\end{align}
Disregarding losses, the collective modes dispersion is identical to
that found by Callaway~\cite{Callaway68}. The quadratic spin
dispersion emerges as ferromagnetism breaks time-reversal symmetry,
and is equal to that of a single minority spin species particle
propagating through a sea of majority spin particles. The dispersion
rises with increasing interaction strength $k_{\text{F}}a$ as the
system becomes stiffer against spin rotation.  Atom loss introduces an
additional energy penalty for fluctuations, consequentially the
dispersion also rises with the loss rate parameter $\tilde{\lambda}$.
To fully expose the influence that atom loss has over the dispersion
it is useful to focus on the instability to a partially polarized
phase which develops at
\begin{align}
 k_{\text{F}}a=\frac{2^{2/3}3}{5}+\sqrt{\frac{2^{1/3}18}{25}-\tilde{\lambda}^{2}}\punc{.}
 \label{eqn:kFaFromDispersion}
\end{align}
Without three-body loss, the fully polarized phase becomes unstable at
$k_{\text{F}}a=2^{5/3}3/5$ in accordance with the prediction of the
mean-field Stoner model. At mean-field level quantum fluctuations
destroy the ferromagnetic state, so \eqnref{eqn:kFaFromDispersion}
matches the boundary Fig.~\ref{fig:LossyFerroIntVar}(a) which
demonstrates how increased loss reduces the required interaction
strength, whereas Fig.~\ref{fig:LossyFerroIntVar}(b) highlighted the
opposite effect when quantum fluctuations are taken into account.
Working at the mean-field level, there is a maximum loss rate,
$\tilde{\lambda}=2^{2/3}3/5$, beyond which the fully polarized state
cannot be formed.

In addition to renormalizing the dispersion, the presence of a loss
interaction also leads to the decay of the spin excitations.  As the
spin wave propagates the atom spins develop a component in the
opposite spin direction incurring atom loss, which in turn damps the
spin wave.  The characteristic inverse timescale of damping, or width,
of a transverse mode can be found from the imaginary component of its
frequency,
\begin{align}
 \Gamma=\frac{q^{2}}{2}\frac{2^{5/3}3\tilde{\lambda}}{5(k_{\text{F}}a)^{2}}\frac{1}{1+\tilde{\lambda}^{2}/(k_{\text{F}}a)^{2}}\punc{.}
\end{align}
We see that due to atom loss the spin-waves become resonances
that are characterized by a momentum independent quality factor
\begin{align}
 Q\equiv\frac{\Omega}{\Gamma}=\frac{5}{2^{5/3}3}\left(\frac{(k_{\text{F}}a)^{2}}{\tilde{\lambda}}+\tilde{\lambda}\right)-\frac{k_{\text{F}}a}{\tilde{\lambda}}\punc{.}
\end{align}
In \figref{fig:LossyFerroIntVar}(a) we highlight the region
$Q<1$, where spin excitations completely lose their integrity.

In experiment, these collective modes can be excited and probed by
spin-dependent Bragg spectroscopy. A variable wavelength optical
lattice potential couples asymmetrically to the spin degrees of
freedom, and thereby excites transverse magnetic fluctuations. The
collective mode response could be studied through dynamical
fluctuations of the cloud spatial distribution as a function of
wavelength, laser amplitude, and detuning.

An experimental handle that could modify the atom loss rate would gift
investigators the ability to fully explore the consequences of atom
loss. Though the atom loss rate cannot be reduced, it sets a base
level that can then be artificially raised. An additional
bosonic~\cite{Incao08} or different fermion~\cite{Wenz09} species
could act as the third body in the atom loss process. The loss rate
will be proportional to the density of this third species, which
can be conveniently controlled. The latter case has already been
investigated with the same lithium species employed in the pioneering
ferromagnetism experiment~\cite{Wenz09,Jo09}.

In this letter we have shown that three-body loss damps quantum
fluctuations. This hinders the transition into the ferromagnetic state
and is consistent with the experimental findings. Furthermore, we have
highlighted signatures of this mechanism in the collective mode
spectrum. A novel phenomenology was developed to probe the
consequences of atom loss, and its generality opens the possibility to
explore transitions from a phase with atom loss to one
without. Candidate systems include the BEC-BCS crossover, p-wave
superfluids, and the boson atom-molecule superfluid
transition~\cite{Radzihovsky04}. This system has no base atom loss,
but adding a third species will activate molecule loss from
the atomic side of the transition. Thus the loss rate can be tuned
from zero upwards and the impact on the phase transition fully
exposed.

We thank Eugene Demler, Andrew Green, Ben Simons, and especially Gyu-Boong Jo and
Wolfgang Ketterle for useful discussions. GJC acknowledges the
financial support of the Royal Commission for the Exhibition of 1851
and the Kreitman Foundation. EA was supported in part by ISF,
US-Israel BSF, and the Minerva foundation.


\begin{thebibliography}{99}

\bibitem{Conduit09}
G.J. Conduit, A.G Green and B.D. Simons. Phys. Rev. Lett. {\bf 103}, 207201 (2009).

\bibitem{Belitz98}
D.~Belitz \emph{et al.}, Phys. Rev. B {\bf 58}, 14155 (1998).
M. Shimizu. Proc. Phys. Soc. {\bf 84}, 397 (1964);
D. Belitz, T.R. Kirkpatrick, and T. Vojta. Phys. Rev. B {\bf 55}, 9452 (1997);
J. Betouras, D. Efremov and A. Chubukov. Phys. Rev. B {\bf 72}, 115112 (2005);
D.V. Efremov, J.J. Betouras and A. Chubukov. Phys. Rev. B {\bf 77}, 220401(R) (2008).

\bibitem{Borzi07}
R. Borzi \emph{et al.}, Science {\bf 315}, 214
(2007); M. Uhlarz, C. Pfleiderer and S.M. Hayden,
Phys. Rev. Lett. {\bf 93}, 256404 (2004).

\bibitem{Coleman08}
I.~Berdnikov, P.~Coleman and S.H.~Simon. Phys. Rev. B {\bf 79}, 224403 (2009).

\bibitem{LeBlanc09}
L.J.~LeBlanc \emph{et al.}, Phys. Rev. A {\bf 80}, 013607 (2009).

\bibitem{Conduit08}
G.J. Conduit and B.D. Simons. Phys. Rev. A {\bf 79}, 053606 (2009).

\bibitem{Petrov03}
D.S.~Petrov. Phys. Rev. A {\bf 67}, 010703(R) (2003).

\bibitem{Jo09}
G.-B.~Jo \emph{et al.}, Science {\bf 325}, 1521 (2009).

\bibitem{Conduit10}
G.J. Conduit and B.D. Simons. Phys. Rev. Lett. {\bf 103}, 200403 (2009).

\bibitem{Zhai09}
H.~Zhai, arXiv:0909.4917.


\bibitem{Syassen09}
N. Syassen \emph{et al.}, Science {\bf 320}, 1329 (2009).

\bibitem{Durr09}
S.~Durr \emph{et al.}, Phys. Rev. A {\bf 79}, 023614 (2009).

\bibitem{Daley09}
A.J.~Daley \emph{et al.}, Phys. Rev. Lett. {\bf 102}, 040402 (2009).

\bibitem{Pathria07}
R.K. Pathria, Statistical Mechanics, Pergamon Press (1996).

\bibitem{Abrikosov58}
A.A. Abrikosov and I.M. Khalatnikov, Soviet Phys. JETP {\bf 6}, 888 (1958);
F. Mohling, Phys. Rev. {\bf 122}, 1062 (1961);
R.A. Duine and A.H. MacDonald, Phys. Rev. Lett. {\bf 95}, 230403 (2005).

\bibitem{Huckans09}
J.H.~Huckans \emph{et al.}, Phys. Rev. Lett. {\bf 102}, 165302 (2009).

\bibitem{Callaway68}
J.~Callaway, Phys. Rev. {\bf 170}, 576 (1968).

\bibitem{Incao08}
J.P.~D'Incao, and B.D.~Esry, arXiv:0508.474.

\bibitem{Wenz09}
A.N.~Wenz \emph{et al.}, Phys. Rev. A {\bf 80}, 040702(R) (2009).

\bibitem{Radzihovsky04}
L.~Radzihovsky, J.~Park, and P.B.~Weichman, Phys. Rev. Lett. {\bf 92}, 160402 (2004)

\end{thebibliography}
\end{document}